\documentstyle[twoside,fleqn,npb,epsfig]{article}

\newcommand{\AmS}{{\protect\the\textfont2
  A\kern-.1667em\lower.5ex\hbox{M}\kern-.125emS}}
\newcommand{\sss}{\scriptscriptstyle}
\title{NLO QCD Predictions for associated $t\bar{t}h$ production 
in Hadronic Collisions}

\author{
S.~Dawson\address{Physics Department, Brookhaven National Laboratory, 
Upton, NY 11973, USA} 
\thanks{This work is supported in part by the U.S. Department of Energy 
under grant DE-AC02-76CH00016.}, 
L.~H.~Orr\address{Department of Physics and Astronomy, 
University of Rochester, Rochester, NY 14627, USA}
\thanks{This work is supported in part by the U.S. Department of Energy 
under grant DE-FG02-91ER40685.}, 
L.~Reina\address{Physics Department, Florida State University, 
Tallahassee, FL 32306-4350, USA}
\thanks{This work is supported in part by the U.S. Department of Energy 
under grant DE-FG02-97ER41022.}  
and D.~Wackeroth\address{Department of Physics, SUNY at Buffalo, 
Buffalo, NY 14260, USA} }

\begin{document}

\begin{abstract}
We present the next-to-leading-order (NLO) QCD corrections to the
inclusive total cross section for the production of a Higgs boson in
association with a top anti-top quark pair within the Standard Model
at the Tevatron and the LHC.
\end{abstract}

\maketitle

\section{Introduction}

The search for the Higgs boson of the Standard Model (SM) is one of
the major tasks of the next generation of high-energy collider
experiments. The direct limit on the SM Higgs boson mass, $M_h$, from
LEP2 searches and the indirect limit from electroweak precision data
strongly suggest the existence of a light Higgs boson, $M_h < 193$~GeV
(95 $\%$ C.L.)~\cite{Grunewald:2002wg}, which may be within the reach
of the Fermilab Tevatron $p\bar{p}$ collider~\cite{Carena:2002es}.
However, since the dominant SM Higgs production channels are plagued
with low event rates and large backgrounds, the Higgs boson search at
the Tevatron requires high luminosity, and all possible production
channels should be considered. The production of a SM Higgs boson in
association with a top-quark pair, $p\bar{p}\to t\bar{t}h$, can play a
role for almost the entire Tevatron discovery
range~\cite{Carena:2002es,Goldstein:ag}. Although the $t\bar{t}h$
event rate is small, the signature of such events is quite spectacular
($W^+W^-b\bar{b}b\bar{b}$).

At the CERN LHC $pp$ collider, the associated $t\bar{t}h$ production
mode will play a crucial role in the 110~GeV$\!\le\!M_h\!\le\!$130~GeV
mass region both for discovery and for precision measurements of the
Higgs boson couplings.  This process will provide a direct measurement
of the top-quark Yukawa coupling and will be instrumental in
determining ratios of Higgs boson couplings in a model independent
way~\cite{Belyaev:2002ua}. Such measurements could help to distinguish
a SM Higgs boson from more complex Higgs sectors, e.g., as predicted
by supersymmetry, and shed light on the details of the generation of
fermion masses.

As for any other hadronic cross section, the next-to-leading-order
(NLO) QCD corrections are expected to be numerically important and are
crucial in reducing the (arbitrary) dependence of the cross sections
on the factorization and renormalization scales.  Here we summarize
the calculation of the NLO QCD predictions for associated $t\bar{t}h$
production at the Tevatron and the LHC.  Results for the Tevatron have
been presented in
Refs.~\cite{Reina:2001sf,Reina:2001bc,Beenakker:2001rj,ditt} and for
the LHC in Refs.~\cite{Beenakker:2001rj,ditt,dlls:2002}.  The results
of the two groups are in very good agreement within the statistical
errors.  \newpage

\section{QCD corrections to $t\bar{t}h$ production}

The inclusive total cross section for $pp$ (or $p\bar{p}$) $\to
t\bar{t}h$ at ${\cal O}(\alpha_s^3)$ can be written as:
\begin{eqnarray}\label{eq:sig}
\lefteqn{\sigma_{\sss NLO}(p\,p\hskip-7pt\hbox{$^{^{(\!-\!)}}$} 
\to t\bar{t}h) =\sum_{ij} \frac{1}{1+\delta_{ij} } \int dx_1 dx_2}  \\
&& \hspace*{-0.1cm}\cdot[{\cal F}_i^p(x_1,\mu) {\cal F}_j^{p(\bar{p})}(x_2,\mu)
{\hat \sigma}^{ij}_{\sss NLO}(\mu) + (1\leftrightarrow 2)]\,\,\,,\nonumber
\end{eqnarray}
where ${\cal F}_i^{p, {\overline p}}$ are the NLO parton distribution
functions for parton $i$ in a proton/antiproton, defined at a generic
factorization scale $\mu_f\!=\!\mu$, and ${\hat \sigma}^{ij}_{\sss
  NLO}$ is the ${\cal O}(\alpha_s^3)$ parton-level total cross section
for incoming partons $i$ and $j$, renormalized at an arbitrary scale
$\mu_r$.  We take $\mu_r\!=\!\mu_f\!=\!\mu$.  The NLO parton-level
total cross section ${\hat \sigma}^{ij}_{\sss NLO}$ consists of the
${\cal O}(\alpha_s^2)$ Born cross section and the ${\cal O}(\alpha_s)$
corrections to the Born cross section, including the effects of mass
factorization.  It contains virtual and real corrections to the
parton-level $t\bar{t}h$ production processes, $q\bar{q}\to t\bar{t}h$
and $gg\to t\bar{t}h$, and the tree-level $(q,\bar{q})g$ initiated
processes, $(q,\bar{q})g\to t\bar{t}h(q,\bar{q})$, which are of the
same order in $\alpha_s$.  The main challenges in the calculation come
from the presence of pentagon diagrams in the virtual corrections with
several massive external and internal particles, and from the
computation of the real part in the presence of infrared
singularities.

\subsection{ Virtual Corrections}
The calculation of the virtual corrections to the $q {\overline q}$
initiated sub-process is described in detail in
Ref.~\cite{Reina:2001bc}.  The calculation of the virtual corrections
to $gg\to t\bar{t}h$ is technically similar.  The basic method is to
reduce each virtual diagram to a sum of scalar integrals, which may
contain both ultraviolet (UV) and infrared (IR) divergences.  The
finite scalar integrals are evaluated by using the method described in
Ref.~\cite{Denner:kt} and cross checked with the numerical package
FF~\cite{olden}.  The scalar integrals that exhibit UV and/or IR
divergences are calculated analytically. Both the UV and IR
divergences are extracted by using dimensional regularization in
$d\!=\!4-2\epsilon$ dimensions. The UV divergences are then removed by
introducing a suitable set of counterterms, as described in detail in
Ref.~\cite{Reina:2001bc} The remaining IR divergences are cancelled by
the analogous singularities in the soft and collinear part of the real
gluon emission cross section.

The most difficult integrals arise from the IR-divergent pentagon
diagrams with several massive particles.  In
Refs.~\cite{Reina:2001sf,Reina:2001bc} we calculated the pentagon
scalar integrals as linear combinations of scalar box integrals using
the method of Ref.~\cite{Bern:1993em}.  For the $gg$ initiated process
we also used the method of Ref.~\cite{Denner:kt} and found perfect
agreement between the results of the two methods.  The virtual
corrections to the $gg$ initiated process have an additional
complication with respect to the $q\bar{q}$ case because of the
presence of pentagon tensor integrals with rank higher than one.
Pentagon tensor integrals can give rise to numerical instabilities due
to the dependence on inverse powers of the Gram determinant.  The Gram
determinant vanishes when two momenta become degenerate, i.e.  at the
boundaries of phase space.  These are spurious divergences, which
cause serious numerical difficulties.  We have used two methods to
overcome this problem and found mutual agreement within the phase
space integration statistical uncertainty:
\begin{itemize}
\item Impose kinematic cuts to avoid the phase space regions where the
  Gram determinant vanishes.  Then apply an extrapolation procedure
  from the numerically safe to the numerically unsafe region.
\item Eliminate all pentagon tensor integrals by cancelling terms in
  the numerator against the propagators wherever possible, after
  interfering the pentagon amplitude with the Born-matrix element. The
  resulting expressions are very large, but numerically stable.
\end{itemize}

\subsection{Real Corrections}

The real corrections are computed using the phase space slicing (PSS)
method, in both the double \cite{Harris:2001sx} and single
\cite{Giele} cutoff approaches. In both approaches the IR region of
the $t\bar{t}h+g$ phase space where the emitted gluon cannot be
resolved is defined as the region where the gluon kinematic
invariants:
\begin{equation}
s_{ig}= 2 p_i\cdot p_g=2E_iE_g(1-\beta_i\cos\theta_{ig})
\end{equation}
become small. Here $p_i$ is the momentum of an external quark or gluon
(with energy $E_i$), $\beta_i\!=\!\sqrt{1-m_i^2/E_i^2}$, $p_g$ is the
final state gluon momentum with energy $E_g$, and $\theta_{ig}$ is the
angle between $\vec{p}_i$ and $\vec{p}_g$.  In the IR region the cross
section is calculated analytically and the resulting IR divergences,
both soft and collinear, are cancelled, after mass factorization,
against the corresponding divergences from the ${\cal O}(\alpha_s)$
virtual corrections.

The single cutoff PSS technique defines the IR region as that where
\begin{equation}
s_{ig}<s_{min}\,\,\,,
\end{equation}
for an arbitrarily small cutoff $s_{min}$.  The two cut-off PSS method
introduces two arbitrary parameters, $\delta_s$ and $\delta_c$, to
separately define the IR soft and IR collinear regions according to:
\begin{eqnarray}
&&E_g<{\delta_s\sqrt{s}\over 2}\,\,\,\,\,\mbox{soft region}\,\,\,,\nonumber\\
&&(1-\cos \theta_{ig})<\delta_c\,\,\,\,\,\mbox{collinear region}\,\,\,.
\end{eqnarray}

In both methods, the real contribution to the NLO cross section is
computed analytically below the cutoffs and numerically above the
cutoffs, and the final result is independent of these arbitrary
parameters. With this respect, it is crucial to study the behavior of
$\sigma_{\sss NLO}$ in a region where the cutoff(s) are small enough
to justify the analytical calculations of the IR divergent
contributions to the real cross section, but not so small as to cause
numerical instabilities.

This is the first application of the single cut-off phase space
slicing approach to a cross section involving more than one massive
particle in the final state. The numerical results of both methods
agree within the statistical errors.  In Ref.~\cite{Beenakker:2001rj},
the dipole subtraction formalism has been used to extract the IR
singularities of the real part.  The agreement between these three
very different treatments of the real IR singularities represents a
powerful check of the corresponding NLO calculations.

\begin{figure}[tb]
\includegraphics[width=16pc]{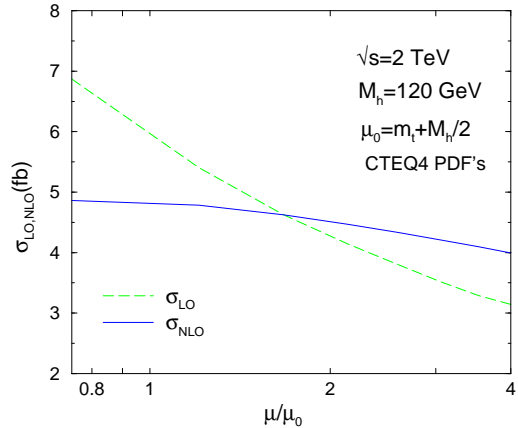}
\vspace*{-1.2cm}
\caption{Dependence of $\sigma_{\sss LO,NLO}(p\bar{p}\to t\bar{t}h)$ 
  on the renormalization/factorization scale $\mu$, at the Tevatron
  ($\sqrt{s}\!=\!2$~TeV), for $M_h\!=\!120$
  GeV~\cite{Reina:2001sf,Reina:2001bc}.}
\label{fg:mudep_tev}
\vspace*{-0.65cm}
\end{figure}

\begin{figure}[tb]
\includegraphics[width=16pc]{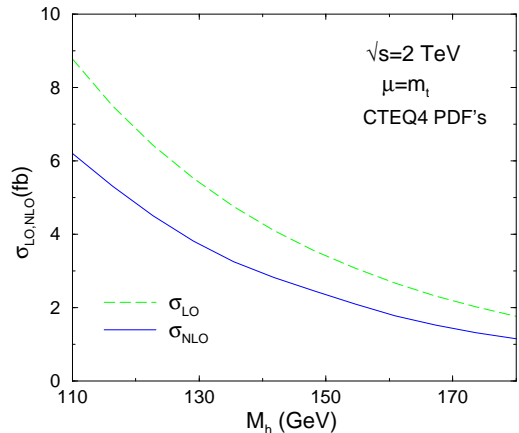}
\vspace*{-1.2cm}
\caption{Dependence of $\sigma_{\sss LO,NLO}(p\bar{p}
  \to t\bar{t}h)$ on $M_h$, at $\sqrt{s}\!=\!2$~TeV, for
  $\mu\!=m_t$~\cite{Reina:2001sf,Reina:2001bc}.}
\label{fg:mhdep_tev}
\vspace*{-0.65cm}
\end{figure}

\begin{figure}[tb]
\includegraphics[width=16pc]{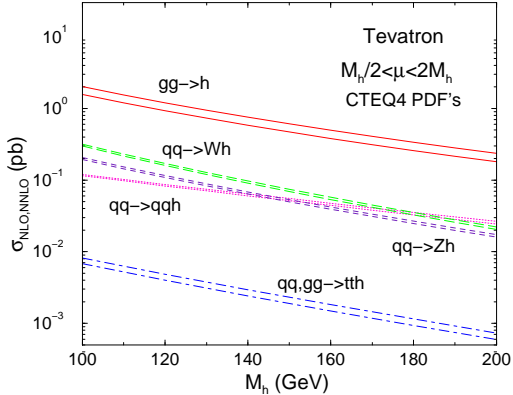}
\vspace*{-1.2cm}
\caption{$\sigma_{\sss NLO,NNLO}$ for SM Higgs production
processes at the Tevatron ($\sqrt{s}\!=\!2$~TeV) 
as a function of $M_h$. 
For $p\bar{p}\to t\bar{t}h$, the renormalization/factorization
scale is varied between $m_t+M_h/2\!<\!\mu\!<\!4(m_t+M_h/2)$.}
\label{fg:NLOtev}
\vspace*{-0.65cm}
\end{figure}

\section{$t\bar{t}h$ production at the Tevatron}

For $p\bar{p}$ collisions at $\sqrt{s}\!=\!2$~TeV, more than $95\%$ of
the tree-level cross section comes from the sub-process $q\bar{q}\to
t\bar{t}h$, while the $gg$ and $(q,\bar{q})g$ initial states are
numerically irrelevant.  Therefore, in
\cite{Reina:2001sf,Reina:2001bc}, when calculating $\sigma_{\sss
  NLO}(p\bar{p}\to t\bar{t}h)$ of Eq.~(\ref{eq:sig}), we only included
the $q\bar{q}\to t\bar{t}h$ channel, summed over all light quark
flavors.  The NLO inclusive total cross section is shown in
Figs.~\ref{fg:mudep_tev} and \ref{fg:mhdep_tev}, as a function of the
renormalization/factorization scale $\mu$ and as a function of the
Higgs boson mass $M_h$, respectively. As expected, the NLO cross
section at the Tevatron shows a drastic reduction in the scale
dependence from the lowest order prediction. The NLO corrections
reduce the total cross section by a factor of $0.7-0.95$ for
$m_t\!<\!\mu\!<\!2m_t$.  Only for $\mu\!>\!2m_t+m_h$ is the NLO cross
section larger than the Born prediction.
 
To conclude, we want to summarize the state of the art of existing NLO
and next-to-NLO (NNLO) calculations for the main SM Higgs boson
production processes at the Tevatron.  To this purpose, in
Fig.~\ref{fg:NLOtev} we show the renormalization/factorization scale
dependence of the inclusive total cross sections at NLO (NNLO in case
of $gg\to h$) for the main SM Higgs production processes. We only omit
$b\bar{b}h$ associated production, since no complete NLO result is yet
available. The renormalization/factorization scales are varied in the
range $M_h/2\!<\mu\!<\!2M_h$ for all processes except $p\bar{p}\to
t\bar{t}h$, where we varied it in the range
$m_t+M_h/2\!<\mu\!<\!4(m_t+M_h/2)$.  The QCD NLO calculations for the
$q\bar{q}\to Wh,Zh$ \cite{Han:1991ia}, $qq\to qqh$ \cite{Han:1992hr}
and $q\bar{q}\to t\bar{t}h$
\cite{Reina:2001sf,Reina:2001bc,Beenakker:2001rj} processes provide
reliable predictions for the inclusive total cross sections at the
Tevatron.  However, the NLO corrections to the $gg\to h$ cross
section~\cite{Dawson:1990zj} are large (up to $\sim 100\%$), and the
NNLO corrections, that recently became available
\cite{Kilgore:2002yw,Anastasiou:2002yz}, are crucial to obtain
reliable predictions~\cite{thanks}.

\section{$t\bar{t}h$ production at the LHC}

The associated production of $t\bar{t}h$ at the LHC with
$\sqrt{s}\!=\!14$~TeV is dominated at the parton level by $gg\to
t\bar{t}h$, and in determining $\sigma_{\sss NLO}(pp\to t\bar{t}h)$,
the ${\cal O}(\alpha_s)$ corrections to the sub-process $gg\to
t\bar{t}h$ provide the largest contribution. However, the other
partonic channels cannot be neglected, and in calculating
$\sigma_{\sss NLO}(pp\to t\bar{t}h)$ of Eq.~(\ref{eq:sig}), we
included all sub-processes, initiated by $gg$, $q\bar{q}$, and
$(q,\bar{q})g$~\cite{dlls:2002}.

\begin{figure}[tb]
\includegraphics[width=16pc]{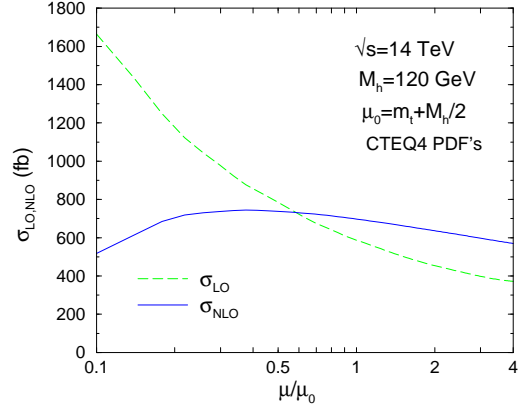}
\vspace*{-1.2cm}
\caption{Dependence of $\sigma_{\sss LO,NLO}(pp\to t
  \bar{t}h)$ on the renormalization/factorization scale $\mu$, at the
  LHC ($\sqrt{s}\!=\!14$~TeV), for
  $M_h\!=\!120$~GeV~\cite{dlls:2002}.}
\label{fg:mudep_lhc}
\vspace*{-0.65cm}
\end{figure}

\begin{figure}[tb]
\includegraphics[width=16pc]{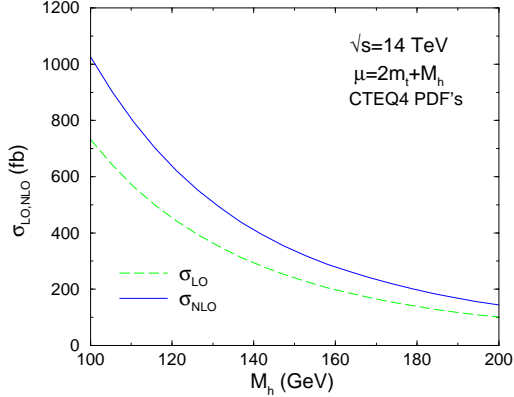}
\vspace*{-1.2cm}
\caption{Dependence of $\sigma_{\sss LO,NLO}(pp \to t\bar{t}h)$ on $M_h$, 
at $\sqrt{s}\!=\!14$~TeV, for $\mu\!=\!2m_t+M_h$~\cite{dlls:2002}.}
\label{fg:mhdep_lhc}
\vspace*{-0.65cm}
\end{figure}

The scale dependence of $\sigma_{\sss NLO}(pp\to t\bar{t}h)$ at
$\sqrt{s}\!=\!14$~TeV is also strongly reduced compared to the LO
result, as can be seen in Fig.~\ref{fg:mudep_lhc}.  The variation of
$\sigma_{\sss NLO}(pp\to t\bar{t}h)$ with the SM Higgs mass is shown
in Fig.~\ref{fg:mhdep_lhc}, for $\mu\!=\!2m_t+M_h$.  At the LHC, the
NLO QCD corrections enhance the LO cross section by a factor of
$1.2\!-\!1.4$ over the entire range of values of the
renormalization/factorization scale shown in Fig.~\ref{fg:mudep_lhc},
and over the relevant Higgs boson mass interval shown in
Fig.~\ref{fg:mhdep_lhc}.

\begin{figure}[tb]
\includegraphics[width=16pc]{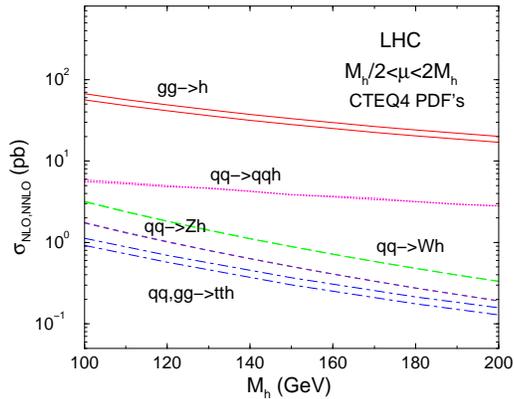}
\vspace*{-1.2cm}
\caption{$\sigma_{\sss NLO,NNLO}$ for SM Higgs production
  processes at the LHC ($\sqrt{s}\!=\!14$~TeV) as a function of $M_h$.
  For $pp\to t\bar{t}h$, the renormalization/factorization scale is
  varied between $m_t+M_h/2\!<\!\mu\!<\!4(m_t+M_h/2)$.}
\label{fg:NLOlhc}
\vspace*{-0.65cm}
\end{figure}

Finally, as we did for the Tevatron in Fig.~\ref{fg:NLOtev}, we
summarize in Fig.~\ref{fg:NLOlhc} the state of the art of existing NLO
and next-to-NLO (NNLO) calculations for the main SM Higgs boson
production processes at the LHC.

\section{Conclusion}

The inclusive cross section for $p\bar{p},pp\to t\bar{t}h$ has been
calculated at NLO of QCD, and the theoretical uncertainty on its
prediction has been drastically reduced to 10-15\% for the
Tevatron and to 15-20\% for the LHC.  

The availability of this NLO result as well as the recent calculation
of the NNLO correction to inclusive Higgs production, allows us to
summarize the state of the art of higher order calculation for Higgs
boson production cross sections at both the Tevatron and the LHC.  The
theoretical predictions for Higgs boson production at the Tevatron and
the LHC are well understood at NLO, and except for the $gg\to h$
process, for which the NNLO corrections are mandatory, the arbitrary
scale dependence at NLO is small.

\section*{Acknowledgments}
We thanks the Authors of Ref.~\cite{Beenakker:2001rj}
  for a detailed comparison of the results.


\begin{thebibliography}{9}

\bibitem{Grunewald:2002wg}
D. Abbaneo {\it et al.} [LEPEWWG], hep-ex/0112021;
M.~W.~Grunewald,
arXiv:hep-ex/0210003.

\bibitem{Carena:2002es}
M.~Carena and H.~E.~Haber,
hep-ph/0208209.

\bibitem{Goldstein:ag}
J.~Goldstein {\it et al.}, 
Int.\ J.\ Mod.\ Phys.\ A {\bf 16S1A}, 308 (2001); 
Phys.\ Rev.\ Lett.\  {\bf 86}, 1694 (2001).

\bibitem{Belyaev:2002ua}
A.~Belyaev and L.~Reina,
JHEP {\bf 0208}, 041 (2002); 
D.~Zeppenfeld,
hep-ph/0203123.

\bibitem{Reina:2001sf}
L.~Reina and S.~Dawson,
Phys.\ Rev.\ Lett.\  {\bf 87}, 201804 (2001).

\bibitem{Reina:2001bc}
L.~Reina, S.~Dawson, and D.~Wackeroth,
Phys.\ Rev.\ D {\bf 65}, 053017 (2002).

\bibitem{Beenakker:2001rj}
W.~Beenakker, S.~Dittmaier, M.~Kr\"amer, B.~Pl\"umper, M.~Spira, and 
P.~M.~Zerwas,
Phys.\ Rev.\ Lett.\  {\bf 87}, 201805 (2001).

\bibitem{ditt}
S.~Dittmaier, these proceedings, hep-ph/0210380.

\bibitem{dlls:2002} S.~Dawson, L.~H.~Orr, L.~Reina, and D.~Wackeroth,
  BNL-HET-02/27, FSU-HEP-2002-1115, UB-HET-02-09.
  
\bibitem{Denner:kt}
A.~Denner,
Fortsch.\ Phys.\  {\bf 41}, 307 (1993).


\bibitem{olden}
G.~J.~van Oldenborgh and J.A. Vermaseren, {\it Z. Phys.} {\bf C46}, 425 (1990).

\bibitem{Bern:1993em}
Z.~Bern, L.~Dixon, and D.~A.~Kosower,
Phys.\ Lett.\ B {\bf 302}, 299 (1993)
[Erratum-ibid.\ B {\bf 318}, 649 (1993)];
Nucl.\ Phys.\ B {\bf 412}, 751 (1994).
%
\bibitem{Harris:2001sx}
For a review see, e.g., B.~W.~Harris and J.~F.~Owens,
Phys.\ Rev.\ D {\bf 65}, 094032 (2002).
%
\bibitem{Giele}
W.~T.~Giele, E.~W.~Glover and D.~A.~Kosower,
Nucl.\ Phys.\ B {\bf 403}, 633 (1993),
S.~Keller and E.~Laenen,
Phys.\ Rev.\ D {\bf 59}, 114004 (1999).

\bibitem{Han:1991ia}
T.~Han and S.~Willenbrock,
Phys.\ Lett.\ B {\bf 273}, 167 (1991).

\bibitem{Han:1992hr}
T.~Han, G.~Valencia, and S.~Willenbrock,
Phys.\ Rev.\ Lett.\  {\bf 69}, 3274 (1992).

\bibitem{Dawson:1990zj}
S.~Dawson,
Nucl.\ Phys.\ B {\bf 359}, 283 (1991);
A.~Djouadi, M.~Spira, and P.~M.~Zerwas,
Phys.\ Lett.\ B {\bf 264}, 440 (1991).

\bibitem{Kilgore:2002yw}
R.~V.~Harlander and W.~B.~Kilgore,
Phys.\ Rev.\ Lett.\  {\bf 88}, 201801 (2002);

\bibitem{Anastasiou:2002yz}
C.~Anastasiou and K.~Melnikov,
hep-ph/0207004.

\bibitem{thanks}
We thank the authors of \cite{Kilgore:2002yw} for providing the NNLO results. 

\end{thebibliography}
\end{document}